\documentclass[runningheads]{llncs}
\usepackage{adjustbox}
\usepackage{graphicx}
\usepackage{amsmath}
\usepackage{amsfonts}
\usepackage{float}
\usepackage{multirow}
\usepackage{subcaption}
\usepackage{xcolor}
\usepackage{breakcites}
\usepackage{hyperref}
\hypersetup{colorlinks=true,citecolor=teal}
\newcommand*\samethanks[1][\value{footnote}]{\footnotemark[#1]}
\begin{document}
\title{Asymmetry Disentanglement Network for Interpretable Acute Ischemic Stroke Infarct Segmentation in Non-Contrast CT Scans}
\titlerunning{Asymmetry Disentanglement Network}
%
\author{Haomiao Ni\inst{1}\thanks{These authors contributed equally to this work.} \and Yuan Xue\inst{2}\samethanks  \and Kelvin Wong\inst{3} \and John Volpi\inst{4} \and \break Stephen T.C. Wong\inst{3} \and  James Z. Wang\inst{1} \and Xiaolei Huang\inst{1}}

\authorrunning{Ni et al.}
\institute{The Pennsylvania State University, University Park, Pennsylvania, USA \and Johns Hopkins University, Baltimore, Maryland, USA 
\and
TT and WF Chao Center for BRAIN \& Houston Methodist Cancer Center, Houston Methodist Hospital, Houston, Texas, USA
\and Eddy Scurlock Comprehensive Stroke Center, Department of Neurology,\break Houston Methodist Hospital, Houston, Texas, USA 
}

\maketitle              
\begin{abstract}
Accurate infarct segmentation in non-contrast CT (NCCT) images is a crucial step toward computer-aided acute ischemic stroke (AIS) assessment. In clinical practice, bilateral symmetric comparison of brain hemispheres is usually used to locate pathological abnormalities. Recent research has explored asymmetries to assist with AIS segmentation. However, most previous symmetry-based work mixed different types of asymmetries when evaluating their contribution to AIS. In this paper, we propose a novel Asymmetry Disentanglement Network (ADN) to automatically separate pathological asymmetries and intrinsic anatomical asymmetries in NCCTs for more effective and interpretable AIS segmentation. ADN first performs asymmetry disentanglement based on input NCCTs, which produces different types of 3D asymmetry maps. Then a synthetic, intrinsic-asymmetry-compensated and pathology-asymmetry-salient NCCT volume is generated and later used as input to a segmentation network. The training of ADN incorporates domain knowledge and adopts a tissue-type aware regularization loss function to encourage clinically-meaningful pathological asymmetry extraction. Coupled with an unsupervised 3D transformation network, ADN achieves state-of-the-art AIS segmentation performance on a public NCCT dataset. In addition to the superior performance, we believe the learned clinically-interpretable asymmetry maps can also provide insights towards a better understanding of AIS assessment. Our code is available at \href{https://github.com/nihaomiao/MICCAI22_ADN}{\color{blue}{https://github.com/nihaomiao/MICCAI22\_ADN}}.
\end{abstract}
\section{Introduction}
\begin{figure}[t]
    \centering
    \begin{subfigure}{0.3\linewidth}
         \centering
         \includegraphics[width=\linewidth]{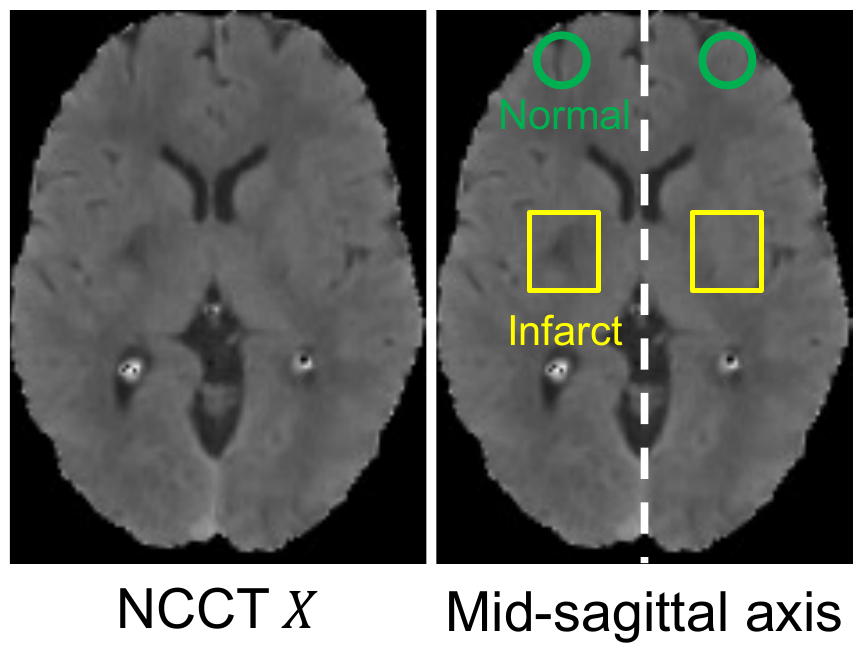}
    \caption{Using bilateral symmetry of the human brain to detect abnormalities.}
    \label{fig:axis}
    \end{subfigure}
    \hfill
    \begin{subfigure}{0.68\linewidth}
         \centering
         \includegraphics[width=\linewidth]{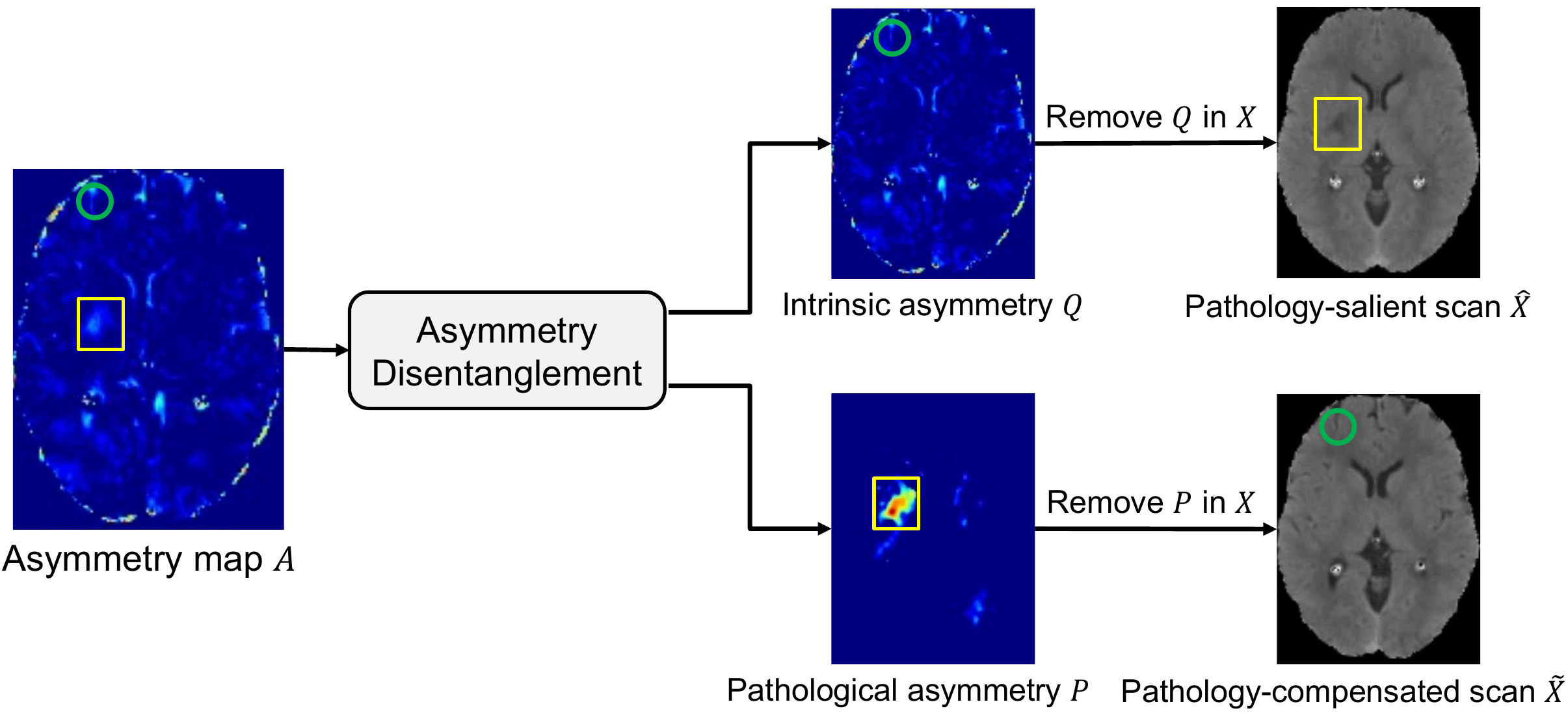}
    \caption{Illustration of the proposed asymmetry disentanglement process. Different types of asymmetries can be removed from the input scan to obtain pathology-salient and pathology-compensated scans. }
    \label{fig:asym_sep}
    \end{subfigure}
    \caption{Illustration of (a) different types of asymmetries in the human brain, and (b) our proposed asymmetry disentanglement framework. Asymmetry maps $A$, $P$, and $Q$ in (b) are normalized for better visualization. (a) and (b) are from the same testing CT scan.}
    \label{fig:illustration}
\end{figure}

Stroke is one of the leading causes of death and disability worldwide~\cite{feigin2021global}. In the United States, about 795,000 people experience a new or recurrent stroke every year, and 87\% of all are acute ischemic strokes (AIS)~\cite{virani2021heart}. 
Non-contrast CT (NCCT) images are routinely used to assess the extent of infarction in AIS patients~\cite{qiu2020machine}. 
For computer-aided AIS estimation in NCCT scans, accurate infarct segmentation is a crucial step.
However, it is challenging to segment AIS infarct in NCCT scans.
First, NCCT is more difficult to process than other medical image modalities such as MRI due to the low signal-to-noise and contrast-to-noise ratios of brain tissues~\cite{kuang2018joint}.
Second, infarct regions can be confounded by normal physiologic changes, and the density and texture variations in involved brain areas may be subtle~\cite{liang2021symmetry}. In clinical practice, to differentiate subtle abnormalities, clinicians often locate suspicious regions by comparing bilateral differences along the mid-sagittal axis (Fig.~\ref{fig:axis}).

Leveraging such prior knowledge that bilateral asymmetries can indicate potential lesions,
recent symmetry-based AIS segmentation approaches~\cite{barman2019determining,kuang2021eis,liang2021symmetry,peter2017quantitative,qiu2020machine,wang2016deep} have shown impressive progress.
Cl\`{e}rigues \textit{et al.}~\cite{clerigues2019acute} exploited brain hemisphere symmetry by inputting CT, CT perfusion images, and their horizontally flipped versions to a 2D-patch-based CNN to segment AIS. 
Kuang \textit{et al.}~\cite{kuang2019automated} proposed a dense multi-path contextual GAN (MPC-GAN) to integrate bilateral intensity difference, infarct location probability, and distance to cerebrospinal fluid (CSF) for AIS segmentation. 

Instead of exploring symmetry at the image level,
Liang \textit{et al.}~\cite{liang2021symmetry} introduced a symmetry-enhanced attention network (SEAN) to segment AIS in NCCT images. The authors first ran a 2D alignment network to transform input images to be bilaterally quasi-symmetric in axial view. Then a symmetry-enhanced attention module was employed to capture both in-axial and cross-axial symmetry at the feature level. 
Though achieving promising results, most existing work simply fused all asymmetries, ignoring the fact that specific asymmetries caused by non-pathological factors cannot reveal clinical findings. For instance, in Fig.~\ref{fig:axis}, the asymmetries between the two yellow boxes indicate left-sided infarction. However, differences between the green circles originate from the natural variation of the brain.
Several recent approaches~\cite{bao2021mdan,chen2020anatomy} explored how to highlight semantically-meaningful pathological asymmetries at the feature level to help identify abnormalities in X-ray or MR images. 
However, such feature-level pathological asymmetry has poor interpretability,
making them less useful in clinical practice.
Also,~\cite{bao2021mdan,chen2020anatomy} both focused on 2D asymmetry analysis, which may fail to make full use of 3D spatial contextual information when applied to CT scans.
Moreover, clinicians usually only look for asymmetry in voxels of the same tissue type, \textit{e.g.}, within the grey matter or within the white matter. Such critical domain knowledge has not been investigated or utilized in the literature.

In this paper, we aim to address the aforementioned weaknesses in existing methods by proposing a novel asymmetry disentanglement network (ADN) to automatically separate pathological and intrinsic anatomical asymmetries in NCCT scans for AIS segmentation with tissue-type awareness. ADN performs asymmetry disentanglement at the image level and achieves high interpretability by directly outputting learned pathological and anatomical 3D asymmetry maps for clinical examination. Furthermore, when asymmetry maps are applied to remove different types of asymmetries from the original NCCT, pathology-salient or pathology-compensated images can be obtained (Fig.~\ref{fig:asym_sep}). Validated by our experiments, performing segmentation on pathology-salient NCCTs shows a noticeable improvement over using original images.
A key novelty of our framework is inspired by the observation that non-pathological asymmetries can be due to intensity differences between different types of tissues, uneven distribution of CSF, or inherent brain anatomical asymmetry. Thus we design a tissue-aware regularization loss function to incorporate tissue-type information into the training of ADN, which further encourages ADN to extract pathologically meaningful asymmetries. Coupled with ADN, an unsupervised 3D transformation network is also proposed to align NCCT scans to obtain the mid-sagittal plane for asymmetry estimation. We conduct extensive experiments on a public NCCT dataset AISD~\cite{liang2021symmetry}, and the results show that our proposed ADN can achieve state-of-the-art performance while successfully disentangling different types of asymmetries in a clinically interpretable way. 

\section{Methodology}
\begin{figure}[t]
    \centering
    \includegraphics[width=\linewidth]{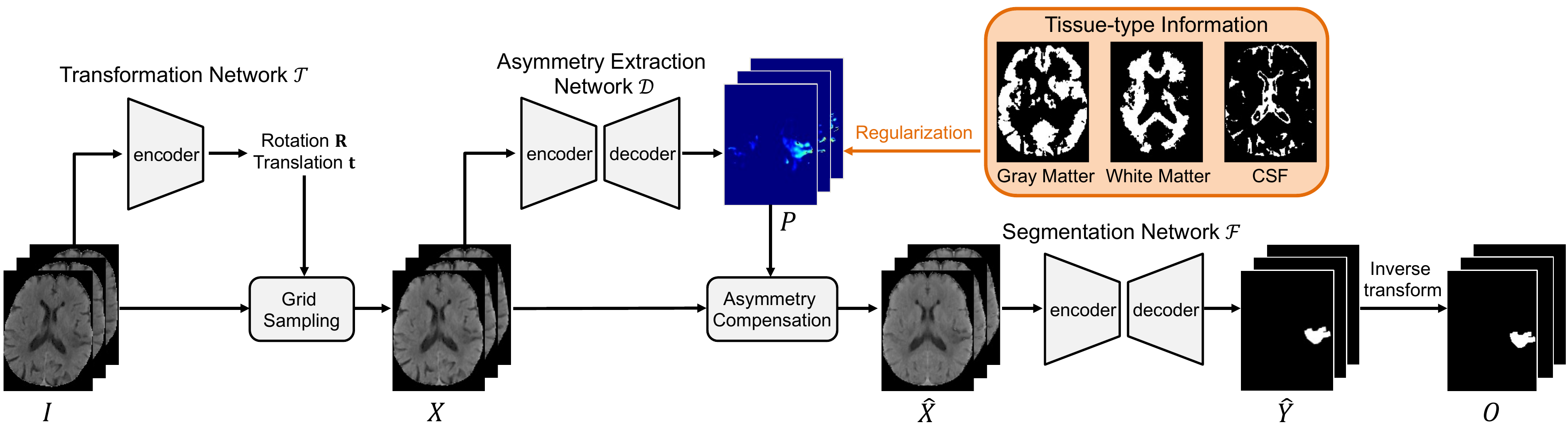}
    \caption{Overview of the asymmetry disentanglement network (ADN). Tissue-type information is generated by \texttt{SPM12}~\cite{ashburner2014spm12} and is only required during training.}
    \label{fig:framework}
\end{figure}
Fig.~\ref{fig:framework} shows the overview of our proposed framework ADN. 
In general, ADN includes three 3D volume-based modules: transformation network $\mathcal{T}$ for input alignment and the subsequent total asymmetry estimation, asymmetry extraction network $\mathcal{D}$ to detect pathological asymmetries and further separate different types of asymmetries, and segmentation network $\mathcal{F}$ for the final AIS segmentation. Network $\mathcal{T}$ is mainly based on a convolutional encoder while both $\mathcal{D}$ and $\mathcal{F}$ adopt an encoder-decoder architecture.
For the input deskulled NCCT scan ${I}$, we first employ an unsupervised 3D transformation network $\mathcal{T}$ to align $I$ and generate a bilaterally quasi-symmetric scan $X$ in which the mid-sagittal plane $S$ 
is in the vertical center of the 3D volume space.
We then feed $X$ into the network $\mathcal{D}$ to extract pathological asymmetry map $P$. $P$ is later used to calculate the intrinsic anatomical asymmetry map and then synthesize an intrinsic-asymmetry-compensated scan $\hat{X}$. We subsequently input $\hat{X}$ to network $\mathcal{F}$ to obtain the segmentation result $\hat{Y}$. We finally apply inverse transformation $\mathcal{T}^{-1}$ to $\hat{Y}$ for getting the final output $O$ corresponding to the input $I$. During the training, we also leverage tissue segmentation results generated from an existing tool to provide extra regularization for network $\mathcal{D}$. The voxel intensities of NCCT images are normalized to be in the range of [0, 1] for network processing and asymmetry computation.

\noindent \textbf{Transformation Network $\mathcal{T}$.} Since the input deskulled NCCT scan $I$ is usually not perfectly centered in the volume space, to make better use of bilateral symmetry, we propose an unsupervised 3D transformation network $\mathcal{T}$ to align the mid-sagittal plane $S$ to be in the vertical center of volume space for subsequent total asymmetry estimation. Different from the 2D alignment network in~\cite{liang2021symmetry}, which may be sensitive to texture variations in some slices, $\mathcal{T}$ utilizes 3D spatial contextual information and outputs parameters $\alpha\in\mathbb{R}^6$, which represents an affine transformation $(\mathbf{R},\mathbf{t})\in\text{SE}(3)$, where $\alpha_\text{1:3}$ are rotation
angles and $\alpha_\text{4:6}$ are translations along $x$, $y$, and $z$ axes.
We then apply the output parameters $\alpha$ to generate bilaterally quasi-symmetric CT image ${X}=\mathcal{T}(I)$ using parameterized sampling grid~\cite{jaderberg2015spatial}. Intuitively, when $S$ is transformed to align with the vertical center plane, $X$ should have the minimum difference from its horizontally flipped version $X^\prime$.
Thus we adopt the following loss function to train the 3D transformation network $\mathcal{T}$:
\begin{equation}\label{eq:loss_t}
  {\displaystyle  L_\mathcal{T} = \left\|{X} - {X}^\prime\right\|_1,}
\end{equation}
where $\left\|\cdot\right\|_1$ is L1 loss. Note that $\mathcal{T}$ computes the 3D alignment in contrast to 2D, and no annotation is required for the training of $\mathcal{T}$.

\noindent \textbf{Asymmetry Disentanglement.}
Based on $X$ and its self-mirrored version ${X}^\prime$, we can compute a total asymmetry map $A$ and further employ an asymmetry extraction network $\mathcal{D}$ to extract pathological asymmetry map $P$ from $A$.
Generally, the ischemic stroke areas appear darker than their contralateral regions in NCCT images (see Fig.~\ref{fig:axis}). Thus voxels that are darker (\textit{i.e.}, having lower intensity) in $X$ than $X^\prime$ are suspicious voxels that could correspond to stroke lesions.
To obtain these asymmetric darker voxels in $X$, we first compute the total asymmetry map $A$ by:
\begin{equation}\label{eq:asym}
{\displaystyle    A_i = \mathrm{max}(X^\prime_i - X_i, 0)\;,}
\end{equation}
where $i$ is any voxel inside $X$. Note that voxels that are darker on one side of the midsagittal plane have positive values in $A$, whereas the value of other voxels is set to zero. We obtain the pathological asymmetry map $P$ through the trained network $\mathcal{D}$, \textit{i.e.}, $P = \mathcal{D}(X)$. Then the map of asymmetry due to intrinsic anatomical asymmetry is $Q = A-P$. Again, in both $P$ and $Q$ asymmetric voxels have positive values and symmetric voxels have zero values. Next, we generate a new image  $\hat{X}$ in which the intrinsic anatomical asymmetry is compensated or neutralized so that there is only pathological asymmetry in $\hat{X}$. 
$\hat{X}$ is defined as:
\begin{equation}\label{eq:enhance}
{\displaystyle    \hat{X} = X + Q = X + A - P\;.}
\end{equation}
Note that although $\hat{X}$ does not exist in reality since actual brain CTs always contain some normal anatomical asymmetry, using this synthetic $\hat{X}$ can make pathological asymmetries more salient and thus make accurate infarct segmentation easier as we will demonstrate in our experiments. Similarly, we can compensate for pathological asymmetry and generate a synthetic image $\Tilde{X}=X+P$ that contains only normal anatomical asymmetry. Our proposed ADN encourages learned anatomical asymmetry and pathological asymmetry to be meaningful via implicit supervision from the final segmentation so that they should not affect pathology information presented in $\hat{X}$ or lead to false positives.
Example asymmetry disentanglement results are shown in Figs.~\ref{fig:asym_sep} and~\ref{fig:asym_show}.
We then input $\hat{X}$ to network $\mathcal{F}$ to obtain segmentation map $\hat{Y}$. The final segmentation $O$ corresponding to the original $I$ can be calculated by applying inverse transformation $\mathcal{T}^{-1}$ to $\hat{Y}$.

\noindent \textbf{Training of Pathological Asymmetry Extraction Network $\mathcal{D}$ and Segmentation Network $\mathcal{F}$.} Intuitively, one can utilize the infarct ground truth map $G$ to train both $\mathcal{D}$ and $\mathcal{F}$ with binary cross-entropy loss:
\begin{equation}\label{eq:seg}
{\displaystyle L_\mathcal{F} = \mathcal{L}_\text{bce}(O, G) = \mathcal{L}_\text{bce}(\mathcal{T}^{-1}(\hat{Y}), G)\;,}
\end{equation}
Since $\hat{Y} = \mathcal{F}(\hat{X})=\mathcal{F}(X+A-\mathcal{D}(X))$, which is related to both $\mathcal{D}$ and $\mathcal{F}$, using the loss ${L}_\mathcal{F}$ can update both $\mathcal{D}$ and $\mathcal{F}$. However, this can lead to a trivial solution $P=\mathcal{D}(X)=A$, such that $\hat{X}$ is equivalent to $X$, downgrading the proposed model to a regular segmentation network. 
To ensure pathological asymmetry map $P$ is clinically meaningful, we 
design a novel tissue-aware regularization loss function to provide extra constraints for the training of $\mathcal{D}$.
This loss function is motivated by
the fact that it is only meaningful to examine pathological asymmetry when the pair of mirroring voxels belong to the same tissue type. Thus $P$ resulting from the network $\mathcal{D}$ should exclude asymmetry in voxels whose mirrored counterparts belong to a different tissue type. 
To utilize such same-tissue constraint, during training, we first employ an off-the-shelf tool, statistical parametric mapping \texttt{SPM12}~\cite{ashburner2014spm12,ashburner1997multimodal} and a CT brain atlas \cite{rorden2012age} to segment GM, WM, and CSF regions from ${X}$; these regions are represented as binary maps $R_\text{GM}$, $R_\text{WM}$ and $R_\text{CSF}$. 
Then we improve these tissue segmentation results by removing the known stroke regions (based on ground truth $G$) from the GM, WM and CSF regions.
The tissue-aware regularization loss for training $\mathcal{D}$ is designed by considering the following loss terms:

\begin{itemize}
\item Tissue-type constraint loss $L_\text{tissue} = \left\|P\cdot (R_\text{GW} + R_\text{CSF})\right\|_1$, which aims to make $P$ exclude asymmetric voxels that belong to different tissue types and regions containing CSF. In particular, $R_\text{GW}$ indicates GM voxels whose mirrored counterparts are WM voxels or WM voxels whose mirrored counterparts are GM voxels; $R_\text{GW}$ can be obtained by computing the intersection of horizontally flipped $R_\text{GM}$ and the original $R_\text{WM}$.

\item\text{Size constraint loss} $L_\text{size} = \left\|\mathrm{mean}(P) - \mathrm{mean}(\mathcal{T}(G))\right\|_1$, which aims to keep $P$ be of similar average size as ground truth.

\item Asymmetry loss $L_\text{asymmetry} = \left\|P\cdot(\mathbf{1}-A)\right\|_1$, which is to constrain all symmetric regions (non-zero in $(\mathbf{1}-A)$) to have zero value in $P$ so that $P$ has non-zero only for asymmetric voxels.

\item Intrinsic asymmetry loss $L_\text{intrinsic} = -\left\|\mathrm{mean}(A-P)\right\|_1$, which is to encourage intrinsic anatomical asymmetries $Q=A-P$ to be as large as possible to contain all non-pathological asymmetries.

\end{itemize}

The final regularization loss function to train $\mathcal{D}$ is calculated by:
\begin{equation}
\label{eq:reg}
{\displaystyle    L_{\mathcal{D}} = L_\text{tissue} + L_\text{size} + L_\text{asymmetry} + L_\text{intrinsic}\;.}
\end{equation}
So the total loss function to jointly train $\mathcal{D}$ and $\mathcal{F}$ is:
\begin{equation}\label{eq:total}
{\displaystyle    L_{\mathcal{DF}} = L_\mathcal{F} + \lambda L_\mathcal{D}\;,}
\end{equation}
where $\lambda$ is a scaling factor. For $L_\mathcal{F}$, we also add an extra generalized Dice loss function~\cite{sudre2017generalised} to alleviate the class imbalance issue. Note that we only use
tissue-type information during training. Using $L_{\mathcal{DF}}$, network $\mathcal{D}$ is trained to help maximize the segmentation accuracy of network $\mathcal{F}$ under the clinical constraints encoded in $L_{\mathcal{D}}$. Thus ADN can learn how to automatically separate different kinds of asymmetries in a clinically interpretable way. 

\section{Experiments}
We validate our proposed method using AISD~\cite{liang2021symmetry}, a public non-contrast CT (NCCT) dataset of acute ischemic stroke (AIS), which includes 345 training scans and 52 testing scans. 
Ischemic lesions are manually contoured on NCCT by a doctor using MRI scans as the reference standard.
The $xyz$ spacing values are various in this dataset, where the $x$- and $y$-spacing values are $0.40\sim2.04$ mm, and $z$-spacing (slice thickness) varies from $3$ to $10$ mm. To ensure that each voxel represents a consistent volume throughout the whole dataset, after skull stripping, we resample all NCCTs to be $1.2\times 1.2\times 5$ $\text{mm}^3$ and reshape them to be $256\times 256\times 40$ using the Python package \texttt{nilearn}~\cite{abraham2014machine}. To evaluate 3D segmentation performance, we employ two volume-based metrics Dice coefficient and Hausdorff distance with 95\% quantile (HD95). The definitions of the metrics can be found in~\cite{taha2015metrics}.

\begin{figure}[t]
    \centering
    \includegraphics[width=0.98\linewidth]{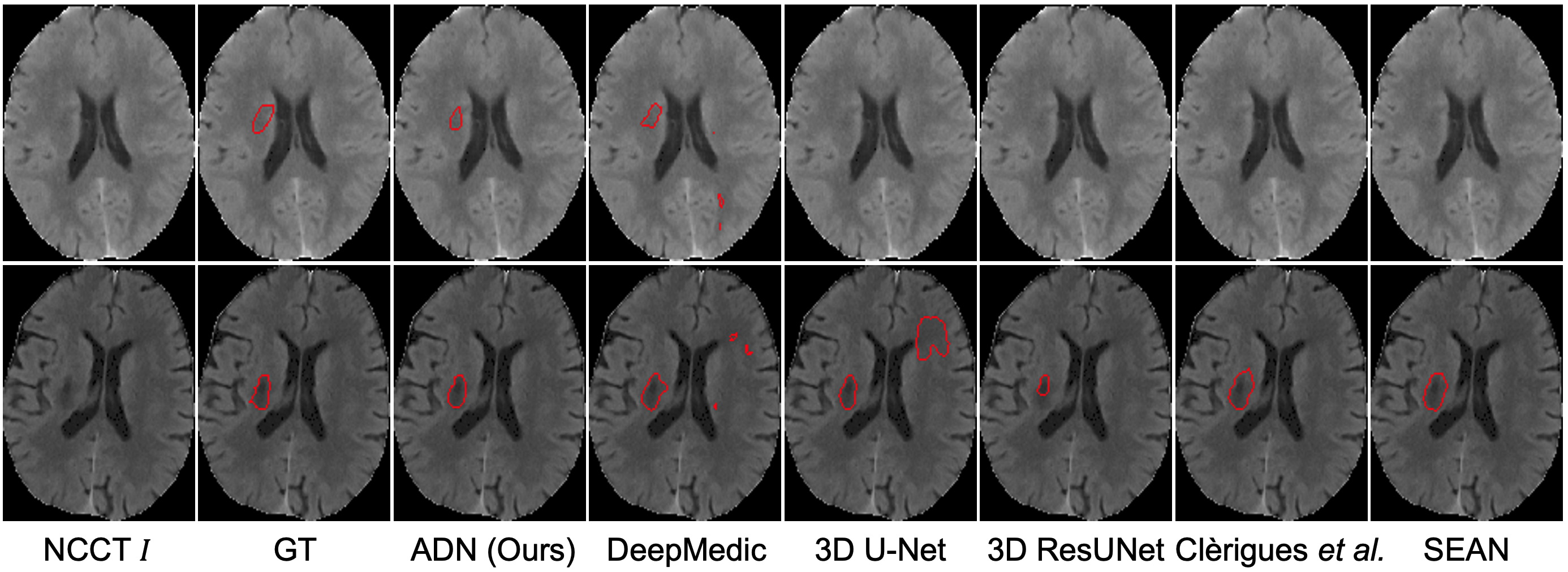}
    \caption{Qualitative comparison of different methods on AISD. Segmented regions are marked by red contours.}
    \label{fig:sota}
\end{figure}

\begin{table}[t]
\centering
\caption{Quantitative comparison of different methods on AISD. ``Aligned'' means whether the transformation is applied to the input CT to make it bilaterally quasi-symmetric.}
\label{tab:sota}
\resizebox{0.5\linewidth}{!}{%
\begin{tabular}{l|c|cc}
\hline
Method         & Aligned & Dice $\uparrow$   & HD95 (mm) $\downarrow$ \\
\hline
DeepMedic~\cite{kamnitsas2017efficient}    & N       & 0.4412 & 58.20 \\
3D U-Net~\cite{cciccek20163d}      & N       & 0.4281 & 42.18 \\
3D ResUNet~\cite{lee2017superhuman}    & N       & 0.4763 & 42.55 \\
Cl\`{e}rigues \textit{et al.}~\cite{clerigues2019acute}        & Y       & 0.5051 & 43.17\\
SEAN ~\cite{liang2021symmetry}         & Y       & 0.5047 & 40.07 \\
\hline
3D ResUNet~\cite{lee2017superhuman} & Y       & 0.4850 & {39.87} \\
ADN w/o $L_\text{tissue}$     & Y       & {0.5090} & {39.66} \\
ADN (Ours)     & Y       & \textbf{0.5245} & \textbf{39.18} \\
\hline
\end{tabular}%
}
\end{table}

\noindent \textbf{Implementation Details.}
Our implementation is based on \texttt{PyTorch}~\cite{paszke2019pytorch} framework.
We implement transformation network $\mathcal{T}$ with four 3D residual blocks~\cite{he2016deep} followed by one fully-connected layer with \texttt{Tanh} activation to predict parameters $\alpha$. 
The range of rotation degrees in $xy$ plane is restricted to be no more than $60 ^{\circ}$ and the translation distance is limited to be no more than half of the image size.
For simplicity, we do not consider rotation and translation of input CT in $z$-axis.
As a general framework, ADN can employ various networks as its backbone. Here we adopt the same architecture as 3D ResUNet~\cite{lee2017superhuman} to implement asymmetry extraction network $\mathcal{D}$ and segmentation network $\mathcal{F}$. 
Due to the GPU memory constraint, we first train unsupervised network $\mathcal{T}$ using Eq.~(\ref{eq:loss_t}) and then fix $\mathcal{T}$ and jointly train $\mathcal{D}$ and $\mathcal{F}$ using Eq.~(\ref{eq:total}). The factor $\lambda$ in Eq.~(\ref{eq:total}) is set to be 10. All networks are trained with \texttt{AdamW} optimizer~\cite{loshchilov2017decoupled} with $(\beta_1, \beta_2)=(0.9, 0.999)$ and $5\times 10^{-4}$ weight decay. The initial learning rates are $1\times10^{-5}$ for network $\mathcal{T}$ and $1\times10^{-3}$ for network $\mathcal{D}$ and $\mathcal{F}$ and we adopt the poly learning rate policy~\cite{chen2017deeplab} with a power of $0.9$. We take the whole CT scan of size $256\times256\times40$ as input and the training batch is set to be 6. We only train models using the training set until convergence. More specifically, we terminate the training once the Dice of the training set remains mostly unchanged for 2 epochs. To help network $\mathcal{D}$ achieve good initialization and facilitate the joint training, at the first 2 epochs, we employ a warm-start strategy by using ground truth $G$ to provide extra supervised cross-entropy loss for training $\mathcal{D}$.

\begin{figure}[t]
    \centering
    \includegraphics[width=0.98\linewidth]{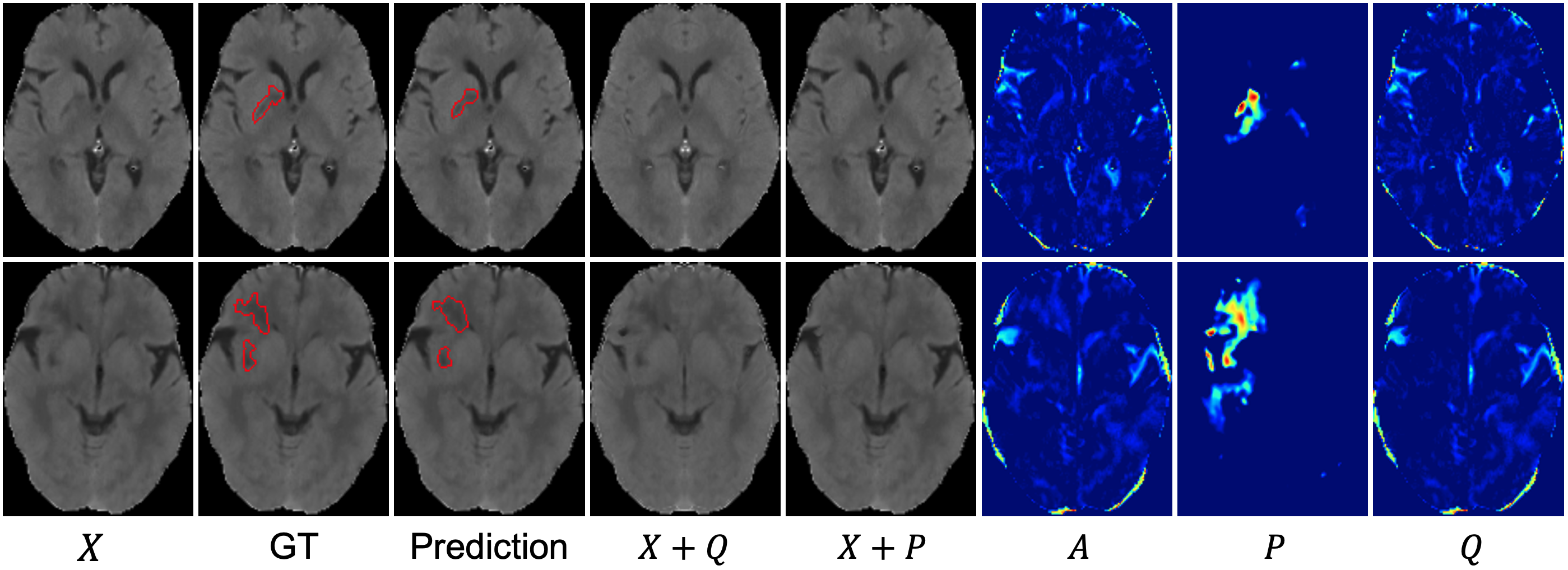}
    \caption{Some asymmetry disentanglement results of our proposed ADN. $X$: aligned NCCT; GT: ground truth; Prediction: AIS segmentation by our method; $\hat{X} = X+Q$: pathology-salient NCCT; $\Tilde{X}=X+P$: pathology-compensated NCCT; $A$: total asymmetry map; $P$: pathological asymmetry map; $Q$: intrinsic asymmetry map. Displayed asymmetry maps are normalized for better visualization.}
    \label{fig:asym_show}
\end{figure}

\noindent \textbf{Result Analysis.}
We compare our proposed ADN with current state-of-the-art methods, including 3D patch-based network DeepMedic~\cite{kamnitsas2017efficient}, volume-based 3D U-Net~\cite{cciccek20163d} and 3D ResUNet~\cite{lee2017superhuman}, and symmetry-aware models Cl\`{e}rigues \textit{et al.}~\cite{clerigues2019acute}
and SEAN~\cite{liang2021symmetry}. The comparison results are shown in Table~\ref{tab:sota} and Fig.~\ref{fig:sota}. The original Cl\`{e}rigues \textit{et al.} is 2D-patch-based and it utilizes symmetry information by concatenating aligned CT patches with their mirrored versions. Here we adapt it to be 3D-volume-based by inputting both aligned CT and its horizontally flipped version to 3D ResUNet. We reimplement SEAN~\cite{liang2021symmetry} according to their paper due to the lack of publicly available implementation. For the other models, we follow publicly available implementations. Note that both Cl\`{e}rigues \textit{et al.} and SEAN are based on aligned CT to better extract bilateral symmetry. 
All the aligned segmentation results will be inversely transformed to correspond to the original NCCT for final comparison. 
As shown in Table~\ref{tab:sota}, ADN has achieved the best Dice and HD95, outperforming all other methods.
We also show some qualitative results in Fig~\ref{fig:sota}.

To further verify the effectiveness of the proposed asymmetry extraction and compensation module, 
we conduct an ablation study by training a 3D ResUNet using aligned CT scans (see the third to the last line in Table~\ref{tab:sota}). The only difference between this baseline and proposed ADN is the input to the segmentation network: 3D ResUNet uses the aligned CT scan ($X$ in Fig. 2) as input while ADN uses the aligned CT scan with intrinsic asymmetry compensated ($\hat{X}$). Results in Table~\ref{tab:sota} demonstrate that using pathology-salient input can achieve a noticeably better Dice coefficient. We also demonstrate the effectiveness of tissue-type constraint loss by ablation study. Compared to the ADN model without using $L_\text{tissue}$, the full ADN model achieves both better Dice and HD95 scores. In addition to the state-of-the-art AIS segmentation performance, ADN also provides clinically interpretable asymmetry maps. We visualize some asymmetry disentanglement results of ADN in Fig.~\ref{fig:asym_show}.

\section{Discussion and Conclusion}
We proposed a novel asymmetry disentanglement network, ADN, to separate different kinds of asymmetries in non-contrast CT (NCCT) images for effective acute ischemic stroke (AIS) infarct segmentation.
Equipped with a clinically-inspired tissue-aware regularization loss function, ADN not only learns pathological and intrinsic anatomical asymmetry maps for clinical interpretation
 but also generates pathology-salient (and intrinsic asymmetry compensated) NCCT images for better AIS detection. 
 We currently focus on NCCT because it is noisier and more challenging than other modalities when dealing with soft tissues. Besides CT, our ADN can be extended to other tasks/modalities that leverage bilateral asymmetries to identify abnormalities. Such tasks include but are not limited to stroke or multiple sclerosis (MS) in brain MRI~\cite{bao2021mdan}, fractures in pelvic X-Rays~\cite{chen2020anatomy}, and infiltration in chest X-Rays~\cite{kim2020learning}.
One limitation of ADN that we observe is that it appears to ignore those bright bleeding spots inside stroke regions. With available annotations, bleeding spots could be detected by a network with inverse intensity change using the same architecture. We plan to explore bleeding spot detection and mitigate their effects in our future work.

\bibliographystyle{splncs04}
\bibliography{paper1796.bib}
\end{document}